\documentclass[usenatbib,usegraphicx,referee]{mn2e}
\usepackage{epsfig}
\def\araa{ARAA}
\def\mnras{MNRAS}
\def\apj{APJ}
\def\aj{AJ}
\def\apjl{APJL}
\def\aap{AAP}
\def\aaps{AAPS}
\def\aapr{AAPR}

\newcommand{\Wt}{\tilde{W}}
\newcommand{\vt}{{\vec \theta}}
\newcommand{\vU}{{\vec{ U}}}

\newcommand{\vr}{{\vec{ r}}}
\def\V{{\cal V}}
\def\N{{\cal N}}
\def\HI{{\rm HI}}
\begin{document}
\title{A Study of ISM of Dwarf Galaxies Using HI Power Spectrum Analysis}
\author[Prasun Dutta, Ayesha Begum, Somnath Bharadwaj and Jayaram
  N. Chengalur]
{Prasun Dutta$^{1}$\thanks{Email:prasun@cts.iitkgp.ernet.in},  
Ayesha Begum$^{2}$\thanks{Email: begum@astro.wisc.edu}, 
Somnath  Bharadwaj$^{1}$\thanks{Email: somnath@cts.iitkgp.ernet.in},
 and Jayaram N. Chengalur$^{3}$\thanks{Email: chengalu@ncra.tifr.res.in}
\\$^{1}$ Department of Physics and Meteorology \&
 Centre for Theoretical Studies, IIT Kharagpur, 721 302  India. 
\\$^{2}$ Department of Astronomy,
University of Wisconsin
475 N. Charter Street Madison, WI 53706, USA 
\\   \ $\&$  Institute of Astronomy, University of Cambridge, Madingley
Road, Cambridge, UK.
\\$^{3}$ National Centre For Radio Astrophysics, Post Bag 3,
Ganeshkhind, Pune 411 007, India.}
\maketitle 

\begin{abstract}
We estimate the power spectrum of HI intensity fluctuations for a
sample of 8 galaxies (7 dwarf and one spiral). The power spectrum can
be fitted to a power law $P_{\rm HI}(U) = A U^{\alpha}$ for 6 of these
galaxies, indicating turbulence is operational. The estimated best fit
value for the slope ranges from $\sim -1.5$ (AND~IV, NGC~628, UGC~4459
and GR~8) to $\sim -2.6$ (DDO~210 and NGC~3741). We interpret this bi-modality
as being due to having effectively 2D turbulence on length scales much
larger than the scale height of the galaxy disk and 3D otherwise.
This allows us to use the estimated slope to set bounds on  the scale
heights of the face-on galaxies in our sample. We also find that 
the power  law slope remains constant as we increase the channel 
thickness for all these galaxies, suggesting that the fluctuations
in HI intensity are due to density fluctuations and not velocity
fluctuations, or that the slope of the velocity structure function is
$\sim 0$. Finally, for the four galaxies with ``2D turbulence'' we
find that the slope $\alpha$  correlates with the star formation
rate per unit area, with larger star formation rates leading to
steeper power laws. Given our small sample size this result needs
to be confirmed with a larger sample.
\end{abstract}


\section{introduction}
\label{sec:intro}
Evidence has been mounting in recent years that turbulence plays an important
role in the physics of the ISM as well as in governing star formation. It is
believed that turbulence is responsible for generating the hierarchy
of structures present across  a range of spatial scales in the ISM
 (e.g. \citealt{ES04I}; \citealt{ES04II}). In such 
models the ISM has a fractal structure and the power spectrum of  intensity
fluctuations is a power law, indicating that there is no preferred
``cloud" size. 

On the observational front, power spectrum analysis of  HI
intensity fluctuations is  an important technique to probe the
structure of the neutral ISM in galaxies (\citealt{L95}; \citealt{OY02}).
The power spectra of the HI intensity fluctuations in our own galaxy, the LMC   
and the SMC all show power  law behavior (\citealt{CD83}; \citealt{GR93};
\citealt{DD00};  \citealt{EK01}; \citealt{SSD99}) which is a
characteristic of a turbulent medium.
   Similarly, \citet{WC99} showed
that the HI distribution in several galaxies  in the
M81 group has a fractal distribution. \citet{KHS82} have estimated the
power spectrum of optical intensity fluctuation for the galaxies M~81,
M~51 and NGC~1365 using digitized images in different bands. The
power spectra  were found to   show both long and short range order. 
Further, \citet{WED05} used the Fourier transform  power spectra of
the V and H$\alpha$ images of a sample of irregular galaxies to show
that the power spectra in optical and H$\alpha$ pass-bands are also
well fit by power laws, indicating that there is no 
characteristic mass or luminosity scale for OB associations and star
complexes.  
\citet{RBDC09} have estimated the power spectrum of the supernovae
remnants Cas~A and Crab.  The power spectrum is found to be
power law, indicating the presence of  magneto-hydrodynamic
turbulence.

Recently \citet{AJS06} have presented a visibility based formalism for
determining the power spectrum  of  HI intensity fluctuations in
galaxies with extremely weak emission. This formalism  is very similar
to that  used for analyzing radio-interferometric observations of the
Cosmic Microwave Background Radiation anisotropies \citep{HO95, WH99,
  HO02, DI04}. This
formalism was first applied to a dwarf
galaxy, DDO~210. Interestingly, the HI power spectrum of this
extremely faint, 
largely quiescent galaxy  was found to be a power law with the same
slope ($-2.8\pm0.4$) as that observed in much brighter galaxies.
We \citep{PASJ07}  have applied the same technique to an external
spiral galaxy NGC~628 (M~74) and found a different slope
$(-1.7\pm0.2)$. 
In this paper we use the same  formalism  to  measure the power   
spectrum of HI intensity fluctuation for a sample of 6 more dwarf
galaxies. For completeness,  
we also include the existing  results for   DDO~210 and NGC~628, and
search for correlations with other galaxy properties, e.g, the star
formation rate (SFR). Any detected correlation would provide insights
into the  nature 
of turbulence  in the ISM. In section 2 we  discuss  the  galaxies 
in our sample. In section 3. we discuss  the power spectrum
estimator used in our work. In section 4, we perform  simulations to
support the  analytical results obtained in section 3. The method of   
analysis is discussed in  section 5. In the last section we present
our results and  conclusions. 

\section{The galaxy Sample}
\label{ref:data}
We briefly discuss some of the relevant   properties of 
the galaxies in our sample. A few  of the galaxy parameters are also 
summarized in Table~\ref{tab:profile} which  contains - 
Row(1): type of the galaxy; Row(2): distance in Mpc; Row(3):  Log[SFR] in M$_{\odot}$yr$^{-1}$  measured from H$\alpha$ emission; Row(4): angular extent of the galaxy calculated from Moment 0 map at HI column density $10^{19}$ atoms~cm$^{-2}$; Row(5): velocity dispersion (km~s$^{-1}$), Row(6): inclination in HI and Row(7): references. 

\begin{table*}
\centering
\begin{tabular}{|l|c|c|c|c|c|c|c|c|c|}
\hline
Galaxies & &DDO~210 & NGC~628 & NGC~3741 & UGC~4459 & GR~8 & AND~IV & KK~230 & KDG~52 \\
\hline \hline
(1) Type  & &Dwarf  & Spiral & Dwarf & Dwarf & Dwarf & Dwarf & Dwarf & Dwarf \\
(2) Distance (Mpc)& &1.0&  8.0   &  3.0  &  3.56 &  2.1  &  6.7 & 1.9   &   3.55 \\
(3) Log[SFR] (M$_{\odot}$yr$^{-1}$) & & $>$-5.42 &-0.1 &-2.47 & -2.04 & -2.46 &  -3.0 & $>$-5.53   &  $>$-5.1 \\
\hline 
(4) Angular extent & &$5'\times4.5'$&$13'\times12'$&$27'\times12'$&$4.5'\times4'$&$4.2'\times4'$&$15'\times8.5'$&$3'\times2'$&$3.5'\times3'$\\
\hline 
(5) $ \sigma_{obs}($km s$^{-1}$) & &6.5(1.0)&8.5(1.5)&$\sim 8.0$&9.0(1.6)&9.0(0.8)&$\sim 8.0$&7.5(0.5)&9.0(1.0)\\
(6) $ i_{HI}$& & 26.0$^{\circ}$ & 6.5$^{\circ}$ & 68.0$^{\circ}$ & 30.0$^{\circ}$ & 27.0$^{\circ}$ & 55.0$^{\circ}$ & 50.0$^{\circ}$ & 23.0$^{\circ}$ \\
(7) References & & 3, 6& 1,8 & 4, 6 & 5,6 &2, 6&7, 6&5, 6&5, 6 \\
\hline
\end{tabular}
\caption{
1- \cite{KB92}, 
2- \cite{AJ03}, 
3- \cite{AJ04}, 
4- \cite{AJK07}, 
5- \cite{AJK06}, 
6- \cite{AJK08}, 
7- Chengalur et al. 2009 (in preparation)
8 - \cite{PR07} 
}
\label{tab:profile}
\end{table*}

\subsection*{DDO~210}
DDO 210 is the faintest ($M_{B}\sim -10.9$) relatively close (at a distance $950\pm50$ kpc; \citealt{L99}) gas-rich member of the Local Group. The HI disc of the galaxy is nearly face-on. On large scales, the HI distribution is not axisymmetric; the integrated HI column density contours are elongated towards the east and south. No H$\alpha$ emission
was detected indicating a lack of on-going star formation in the
galaxy. Distance to this galaxy is estimated to be 950$\pm$50 kpc in 
\cite{L99}. 

\subsection*{NGC~628}
NGC~628 (M74) is a nearly face-on  SA(s)c spiral galaxy with an
inclination angle in the range  $6^{\circ}$ to $ 13^{\circ}$ \citep{KB92}.  
It has a very large HI disk extending out to more than 3 times the
Holmberg diameter.  \citet{EE06} have found a
scale-free size and luminosity  distribution of  star forming regions in
this galaxy, indicating turbulence to be functional here.  
The distance to this galaxy is uncertain with previous estimates
ranging from $6.5 \, {\rm Mpc}$ to $ 10 \, {\rm Mpc}$. \cite{SK96}
estimated a distance of $7.8\pm 0.9$ Mpc from  the brightest blue star
in the galaxy. This  distance estimate matches 
with an independent photometric distance estimate by \citet{SD96}. 
In a recent study \citet{VI04} inferred the  distance to be $6.7 \pm  4.5 $ 
Mpc by applying  the expanding  photosphere method to the hyper novae SN2002ap.
In this paper, we adopt the photometric distance of 8.0 Mpc for this galaxy. 

\subsection*{NGC~3741}
NGC~3741 is a nearby dwarf irregular galaxy ($M_{B}\sim -13.13$) with
a gas disk that extends to $\sim 8.8$ times the Holmberg radius
(\citealt{AJK05, AJK07}). The galaxy is fairly edge-on with
kinematical  inclination varying from $\sim 58^{\circ}-70^{\circ}$.
NGC~3741  appears to have a HI bar  and is very dark matter dominated with a dark to luminous mass ratio of 
$\sim 149$. Further, this galaxy is undergoing significant star formation in the center. An interplay between the neutral ISM and star formation in this
galaxy is studied in detail in \citet{AJK07}. 
 The rotation curve flattens beyond 
$300^{\prime \prime}$ to a value $\sim 50\ {\rm  km \, s}^{-1}$ \citep{GPU07}. They find  that the ISM in this galaxy shows radial
 motion of   $\sim 5 -  13 \, {\rm km \,  s}^{-1}$.  
 \citet{KKH04} have estimated the  distance 
to this galaxy as $3.0\pm0.3$ Mpc using the tip of
the red giant branch (TRGB) method. 

\subsection*{UGC~4459}
The faint dwarf ($M_{B}\sim-13.37$) galaxy UGC~4459 is a member of the 
M~81 group of galaxies. It is fairly isolated  
from its nearest neighbor UGC~4483 at a projected distance of
$3.6^{\circ}~(\sim 223$ kpc) and a velocity difference $135\ $km 
s$^{-1}$.  
UGC~4459 is a relatively metal poor galaxy, with $12\ +\ \log($O/H$)\sim
7.62$ \citep{KO00}. The optical appearance 
of  this galaxy is dominated by bright blue clumps, which emits
copious amount of H$\alpha$, indicating high star formation.  
The velocity field of UGC~4459 shows a large scale gradient across the
galaxy with an average of $\sim 4.5$ km s$^{-1}$ kpc$^{-1}$, 
though this gradient is not consistent with that expected from a
systematic rotating disk. 
This galaxy has  a TRGB distance of $3.56$ Mpc \citep{KKH04}.

\subsection*{GR~8}
This  is a faint ($M_{B}\sim -12.1$) dwarf irregular galaxy with very
unusual HI kinematics (\citet{AJ03}). 
The HI distribution in the galaxy is very clumpy, and  shows 
substantial diffuse, extended gas.   
The high density HI clumps in the galaxy are associated with 
optical knots.  In optical it has  
a patchy appearance with the emission dominated by  bright blue
knots  which are sites of active star formation \citep{H67}. 
Both radial and circular   motions are  present in this
galaxy.  It also possesses a  faint extended emission in H$\alpha$. 
 The distance to this galaxy is estimated to be  $2.10$ Mpc (\citealt{KKH04}).

\subsection*{AND~IV}
This  is a dwarf irregular galaxy with a moderate surface brightness 
($\bar{\mu}_{V} \sim 24$) and a very blue colour ($V-I \leq 0.6$)
\citep{FGW00}. 
It  is  at a projected distance of $40^{'}$  from the
center of M~31 \citep{M06} and  has  a very low ongoing
SFR of $\sim 0.001 M_{\odot} $yr$^{-1}$.  Its   
 HI disk  extends to  $\sim 6$ times the Holmberg 
diameter and also shows large scale, purely gaseous spiral arms
Chengalur et al. 2008 (in preparation). The distance to this galaxy is
estimated to be $6.7\pm 1.5$ Mpc \citep{FGW00}.  

\subsection*{KK~230}
KK~230, the faintest($M_{B}\sim -9.55$) dwarf irregular galaxy in our sample, lies at the periphery  of the Canes Venatici~I cloud of 
galaxies \citep{KKH04}. The velocity field  shows a gradient in
the direction roughly perpendicular to the HI and 
optical major axis with a magnitude of $\sim 6\ $km s$^{-1}$
kpc$^{-1}$ \citep{AJK06}. There is no measurable ongoing star formation  
in this galaxy as inferred from the absence of any detectable 
H$\alpha$ emission. \citet{KKH04} have found a tidal index of  
$-1.0$ indicating the galaxy to be  fairly isolated. They have
estimated the TRGB distance  to be  $1.9$ Mpc.

\subsection*{KDG~52}
KDG~52 (also called M~81DwA), another member of the M81 group, is a
faint dwarf  galaxy ($M_{B}\sim-11.49$)  with  a clumpy  HI 
distribution in  a broken ring surrounding the optical
emission \citep{AJK06}.  
The HI hole is not exactly centered around the optical emission.
This galaxy does not have any detectable ongoing star formation. 
The distance to this galaxy is estimated to be $3.55$ Mpc \citep{KKH04}
derived from the TRGB  method.

\section{ A visibility based power spectrum estimator}
\label{sec:method}
\begin{figure}
\begin{center}
\epsfig{file=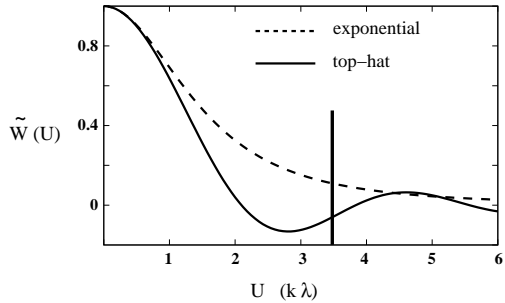,width=2.6in, angle=0}
\end{center}
\caption{$\Wt(\vU)$ for top-hat and exponential window functions for
  $\theta_{0} = 1'$. The vertical line marks   $\theta_{0}^{-1}$. Note
  that the FWHM of the two window functions are nearly the same and both
  window functions have a very small value for  $U >  \theta_{0}^{-1}.$} 
\label{fig:wind}
\end{figure}
The specific intensity of  HI emission from a  galaxy 
can  be modeled as 
\begin{equation}
I_{\nu}(\vt) \ =\ W_{\nu}(\vt)\left[ \bar{I}_{\nu} \ +\ \delta
  I_{\nu}(\vt) \right ]\,.
\label{eq:a1}
\end{equation}
Here $\vt$ is the angle  on the sky  measured in radians 
from the center of the galaxy. We assume that the
galaxy subtends a  small angle so that $\vt$ may be treated as a two
dimensional (2D) planar vector on the sky. The HI specific intensity
is modeled as  the sum of a  smooth component and a 
fluctuating component.  Typically, $I_{\nu}(\vt)$ is maximum
at the center and declines  with increasing $\theta$.
We model this through a  window function   $W_\nu(\vt)$ 
which is defined so that $W_\nu(0)=1$ at the center and has values $1
\ge W_\nu(\vt) \ge 0$ elsewhere. This multiplied by $\bar{I}_\nu$
gives the  smooth component of the specific intensity. 
For a face-on  galaxy, the window function $W_\nu(\vt)$  corresponds
to the galaxy's  radial profile.

Consider a galaxy of angular radius  $\theta_0$. 
In our analysis we  consider two different models for the window
function of such a galaxy. In the ``top-hat'' model it is assumed
that the specific 
intensity has a constant value within a  circular disk of
radius $\theta_0$, and it abruptly falls to $0$ outside. 
The window function, in this case,  is a Heaviside step function
\begin{equation}
W_\nu(\vt)=\Theta(\theta_0-\theta)
\end{equation}
 In the ``exponential'' model the window function has the form 
\begin{equation}
W_{\nu}(\vt) =\exp\left(-\frac{\sqrt{12} \theta}{ \theta_0}\right)\,.
\label{eq:expw}
\end{equation}
where it falls exponentially away from the center. 

It is also possible to use $W_{\nu}(\vt)$ to  define a normalized window
function $W^N_{\nu}(\vt)$ such that $\int d^2 \theta
\, W^N_{\nu}(\vt)=1$. The second moment of $\theta$ defined as $\int
d^2 \theta\, \theta^2 \, W^N_{\nu}(\vt)$ provides a good estimate of
the angular extent of the window function. Here we have used the
condition that 
this should have the same value $\theta^2_0/2$ for all models 
of the window function, to 
determine the width of the exponential window function in terms of
$\theta_0$.  
Throughout we have assumed $\theta_0 \ll 1\ {\rm radian}$.

We express the fluctuating component of the specific intensity as 
$W_\nu(\vt) \, \delta I_\nu(\vt)$. Here $\delta   I_\nu(\vt)$ is a
stochastic fluctuation which is assumed to be statistically homogeneous
and isotropic. 
The HI emission traces these fluctuations modulated by
the window function which quantifies the large-scale HI distribution. 
In this paper we would like to use radio-interferometric observations
to  quantify the statistical properties of the fluctuations $ \delta
I_\nu(\vt)$. These fluctuations  are believed to be the outcome of
turbulence  in the inter-stellar medium. 

The visibility $\V_\nu(\vU)$ recorded in radio-interferometric
observations is the Fourier transform of the product of the antenna
primary beam pattern $A_{\nu} (\vt)$ and the specific intensity
distribution of the galaxy. 
\begin{equation}
\V_{\nu} (\vU) \ =\ \int d \vt \ e^{-i 2 \pi \vU . \vt} A_{\nu} (\vt) \,
I_{\nu }(\vt)
\end{equation}
Here $\vU$ refers to a baseline,  the antenna separation measured in
units of the observing  wavelength $\lambda$. It is common practice to
express $\vU$ in units of kilo wavelength  (k$\lambda$). Throughout
this paper we follow the usual radio inteferometric convention and
express $\vU$ in units of kilo wavelengths (k$\lambda$).

The angular extent of the galaxies that we consider here is
much smaller than the primary beam, and the effect of the primary beam
may be ignored. We then have 
\begin{equation}
\V_\nu(\vU)=\Wt(\vU) \bar{I}_\nu + \Wt(\vU)  \otimes
\tilde{\delta I_{\nu}}(\vU)  + \N_\nu(\vU)
\label{eq:vis2}
\end{equation}
where  the tilde $\tilde{\,}$ denotes the Fourier transform of the
corresponding quantity and $\otimes$ denotes a convolution. 
In addition to the signal,  each visibility
also contains a system noise contribution $\N_\nu(\vU)$ which we have
introduced in  
eq. (\ref{eq:vis2}). The noise in each visibility is a Gaussian random
variable and the noise in the visibilities at two different baselines
$\vU$ and $\vU^{'}$ is uncorrelated. 

The Fourier transform of the normalized window functions  are 
\begin{equation}
\tilde{W}_\nu(\vU)  =  2 ~\frac{J_{1}(2 \pi \theta_0 U)}
{2 \pi \theta_0 U}
\end{equation}
and 
\begin{equation}
\tilde{W}_\nu(\vU)  = \frac{1}{\left[ 1 + \pi^{2} \theta_{0}^{2}U^{2}/3\right]^{3/2}} 
\label{eq:gausw}
\end{equation}
for the top-hat and exponential models respectively. Here $J_1(x)$ is the
Bessel function of order $1$. These functions,   shown in
Figure~\ref{fig:wind}, have the property that they peak around $U=0$
and  fall off rapidly  for $U \gg  \theta_0^{-1}$. This is a generic
property of the window function, not restricted to just these two
models. At baselines $U \gg  \theta_0^{-1}$ we may safely neglect the
first term in eq. (\ref{eq:vis2}) whereby 

\begin{equation}
\V_\nu(\vU)= \Wt(\vU)  \otimes
\tilde{\delta I_{\nu}}(\vU)  + \N_\nu(\vU)
\label{eq:vis3}
\end{equation}
 
We use the power spectrum of HI intensity fluctuations $P_{\HI}(U)$
defined as
\begin{equation}
\langle \tilde{\delta I_\nu}(\vU) \tilde{\delta I_\nu}^{*}(\vU')
\rangle =\delta_{D}^{2}(\vU-\vU') \, P_{\HI}(U)
\end{equation}
to quantify the statistical properties of the intensity
fluctuations. Here, $\delta_{D}^{2}(\vU-\vU')$ is a two dimensional
Dirac delta function. The angular brackets here denote an ensemble
average over different realizations of the stochastic fluctuation. In
reality, it is not possible to evaluate this ensemble average because
a  galaxy presents us with only a single realization. In practice we 
evaluate an angular average over different directions of $\vU$. This
is expected to provide an estimate of the ensemble average for a
statistically isotropic fluctuation. 

The square of the visibilities can, in principle, be used to estimate
$P_{\HI}(U)$ 
\begin{equation}
\langle\ \V_{\nu}(\vU)\V^{*}_{\nu}(\vU)\ \rangle \ = \left
|\Wt_{\nu}(\vU)  \right|^{2} \otimes   \ P_{HI}(\vU) + \langle \mid
\N_\nu (\vU) \mid^2 \rangle \,
\label{eq:noise}
\end{equation}
and this   has been used  in several earlier studies 
\citep{CD83,GR93,L95}.  This  technique has  limited utility 
to observations where the HI signal in each visibility exceeds the
noise.  This is because the last term  $\langle \mid\N_\nu (\vU)
\mid^2 \rangle$, which  is  the noise variance  introduces a
positive bias in estimating the 
power spectrum. The  noise bias can be orders of magnitude larger than
the power spectrum for the  faint external galaxies considered here.  
In principle, the   noise bias may be separately estimated using  
line-free channels and subtracted. In practice this is extremely
difficult owing to uncertainties in  the bandpass response which
restrict the noise statistics to be estimated with the required
accuracy.  Attempts  to subtract out the noise-bias have shown  that
it is not possible to do this at the level of accuracy required to
detect the HI power spectrum \citep{begumt}. 

\begin{figure}
\begin{center}
\epsfig{file=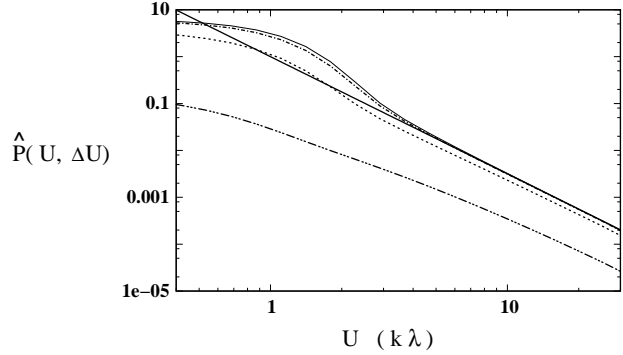,width=3.2in, angle=0}
\end{center}
\caption{This shows the convolution   (eqn. \ref{eq:corr})
of a power law $P_{\HI}(U)=U^{-2.5}$  with the product of two exponential
window functions (eqn. \ref{eq:gausw})  with $\theta_0=1'$ ($(\pi
\theta_0)^{-1} \approx 1\, {\rm k}\lambda$). The bold solid curve shows the
original power law and  the thin solid curve shows $P_{\HI}(\vU,\Delta
\vU)$ for $\Delta \vU=0$. The dashed curves are for$\mid \Delta
\vU\mid =0.2, 1.0 $  and $3.0 \, {\rm k} \lambda$ respectively from top
to bottom. The direction of $\Delta \vU$ is parallel to that of
$\vU$. Note that $P_{\HI}(\vU,0.2) \approx P_{\HI}(\vU,0)$ and 
the correlation between two different visibilities is 
small for $\mid \Delta \vU \mid > (\pi \theta_0)^{-1}$.}
\label{fig:conv2}
\end{figure}

The  problem of noise bias can be avoided by correlating  visibilities
at two different baselines for which  the noise is expected to be
uncorrelated. We define the power spectrum estimator 
\begin{eqnarray}
\hat{\rm P}_{\rm HI}(\vU, \Delta \vU) &=&
\langle\ \V_{\nu}(\vU)\V^{*}_{\nu}(\vU+\Delta \vU)\ \rangle 
\nonumber  \\
&=& \int d^2 U' \, W_{\nu}(\vU-\vU') \, W^*_{\nu}(\vU+\Delta \vU-\vU')
\      P_{\HI}(\vU')  
\label{eq:corr}
\end{eqnarray}
Since $\tilde W (\vU)$ falls  off rapidly for $U \gg
\theta_{0}^{-1}$ (Figure ~\ref{fig:wind}), the 
window  functions $W_{\nu}(\vU-\vU')$ and $W^*_{\nu}(\vU+\Delta
\vU-\vU')$  in eqn.~(\ref{eq:corr})  have a substantial overlap only  
if $|\Delta \vU| < (\pi \theta_{0})^{-1}$.  Visibilities at two  different
baselines  will be correlated only if  $|\Delta \vU| <
(\pi \theta_{0})^{-1}$, and not beyond (Figure~\ref{fig:conv2}). In  our
analysis we restrict the 
difference in baselines to  $|\Delta \vU| \ll  (\pi \theta_{0})^{-1}$
so that  
$\tilde W_\nu(\vU + \Delta \vU-\vU^{'}) \approx \tilde
W_\nu(\vU-\vU^{'})$ and   the estimator $\hat{\rm P}_{\rm
  HI}(\vU, \Delta \vU)$ no  longer depends on $\Delta \vU$
(Figure~\ref{fig:conv2}). We then
use   the visibility correlation estimator 
\begin{eqnarray}
\hat{\rm P}_{\rm HI}(\vU) &=&
\langle\ \V_{\nu}(\vU)\V^{*}_{\nu}(\vU+\Delta \vU)\ \rangle 
\nonumber  \\
&=& \int d^2 U' \, \mid W_{\nu}(\vU-\vU') \mid^2 \,
P_{\HI}(\vU')  \,.
\label{eq:corrn}
\end{eqnarray}
The measured visibility correlation $\hat{\rm  P}_{\rm HI}(\vU)$
will, in general, be  complex. The real part is the power spectrum of
HI intensity fluctuations convolved with the square of the window
function. A further simplification is possible at large baselines 
 $U \gg \theta_{0}^{-1}$, provided $\mid \tilde{W}_\nu(U)\mid^2 $  decays
much faster than the variations in ${\rm P_{\HI}}(U)$. We then have  
\begin{equation}
\hat{\rm P}_{\rm HI}(\vU)= C \,  P_{\HI} (\vU) 
\label{eq:corra}
\end{equation}
where $C=\int  \mid \tilde W_\nu(U) \mid^2 ~d^2  U$ is a constant. 

\begin{figure}
\begin{center}
\epsfig{file=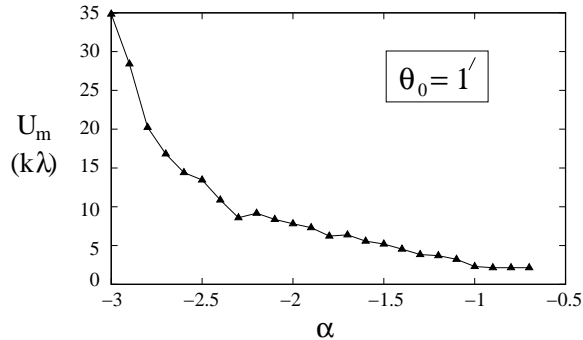,width=3in, angle=0}
\end{center}
\caption{This shows how $U_m$ changes with $\alpha$ for an exponential
window function with $\theta_0=1'$. Note that the discontinuities seen
in the plot appear to be genuine features and not numerical artifacts,
though the cause of these features is not clear  at present.}  
\label{fig:Um}
\end{figure}

\begin{figure}
\begin{center}
\epsfig{file=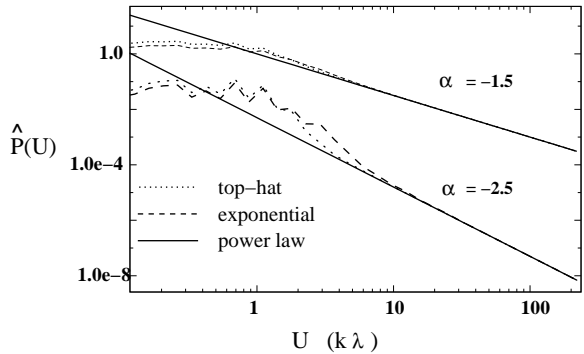,width=3in, angle=0}
\end{center}
\caption{Effect of window function modifies the power spectrum
  differently for different power law exponents. } 
\label{fig:conv}
\end{figure}

We use the real part of the estimator $\hat{\rm P}_{\HI}(\vU)$ to
estimate the power spectrum $P_{\HI}(U)$.  Our interpretation is
restricted to the $U$ range $U \gg 
(\pi \theta_0)^{-1}$ where the convolution  in
eq. (\ref{eq:corrn}) does not affect the shape of the power spectrum
and eq. (\ref{eq:corra}) is a valid approximation. In order to
estimate  the $U$ range where this approximation is valid we have
numerically evaluated eq. (\ref{eq:corr})
assuming $P_{\HI}(U)$ to be a power law $P_{\HI}(U)=A U^{\alpha}$.
Figure \ref{fig:conv} shows the results for both the top-hat and the
exponential models with $\theta_0=1'$ ($\theta_0^{-1}=3.4 \, {\rm
  k}\lambda$) and $\alpha=-1.5$ and $-2.5$ which roughly spans the
range of slopes that we encounter in our  galaxy sample. 
Using $U_m$ to denote the value (in ${\rm k} \lambda$) where 
the deviation from the original power law is $10\%$, we find that 
 for the top-hat and exponential models respectively $U_m$ has values
 $(3.1,5.3)$ for $\alpha=-1.5$ and  $(4.8, 13.0)$ for $\alpha=-2.5$.
Note that  $U_{m}$  depends on two parameters, namely $\alpha$ and
$\theta_{0}$.  
 Figure~\ref{fig:Um} shows how $U_{m}$ changes with $\alpha$ for the
 exponential model with $\theta_0=1'$.  For other  values of
 $\theta_{0}$  we  scale the value   of $U_m$ in  Figure~\ref{fig:Um}
 using  $U_m \propto \theta_0^{-1}$. 
  For a given power law index, the estimator $\hat{\rm P}_{\HI}(\vU)$
  gives a direct estimate of the power  spectrum   for $U \ge U_m$.

 The estimator $\hat{\rm P}_{\HI}(\vU)$ also has a  small  imaginary
 part that arises  from the HI power spectrum  because  the
 assumption that     
$\tilde W_\nu(\vU + \Delta \vU - \vU^{'}) \approx \tilde W_\nu(\vU -
\vU^{'})$  is not strictly valid.  We use the requirement that the  
imaginary part of  $\hat{\rm P}_{\rm HI}(\vU)$ should be small 
 compared to the real part 
as a  self-consistency  check to determine the range  of validity of
our formalism.   

The  real and imaginary parts  of the 
measured value of the estimator $\hat{\rm P}_{\HI}(\vU)$ both have
uncertainties  arising from
 (1.) the sample variance and (2.) the system
noise. The sample variance is $[\hat{\rm P}_{\HI}(\vU)/ \sqrt{N_{E}}]^2$,
where $N_E$ is the number of independent estimates. We assume that in
the u-v plane the 
 HI signal is uncorrelated beyond $1.5$ times the FWHM of
 $|\tilde{W}_{\nu}^{2}(U)|^{2}$. We use this  to determine $N_E$ from the
 u-v coverage of 
 our observations. The system noise variance is $\sigma^4/N_P$ where 
$\sigma^2$ is the variance of the individual
visibilities in the data and   $N_P$ is  the number of visibility
pairs that contribute to each  visibility correlation. We add both
these contributions to determine the $1-\sigma$ error-bars.  The
reader is referred to Section 3.2 of \citet{SS08} for further
details of the error estimation.  
\section{Simulation}
\label{sec:simulation}
\begin{figure}
\begin{center}
\epsfig{file=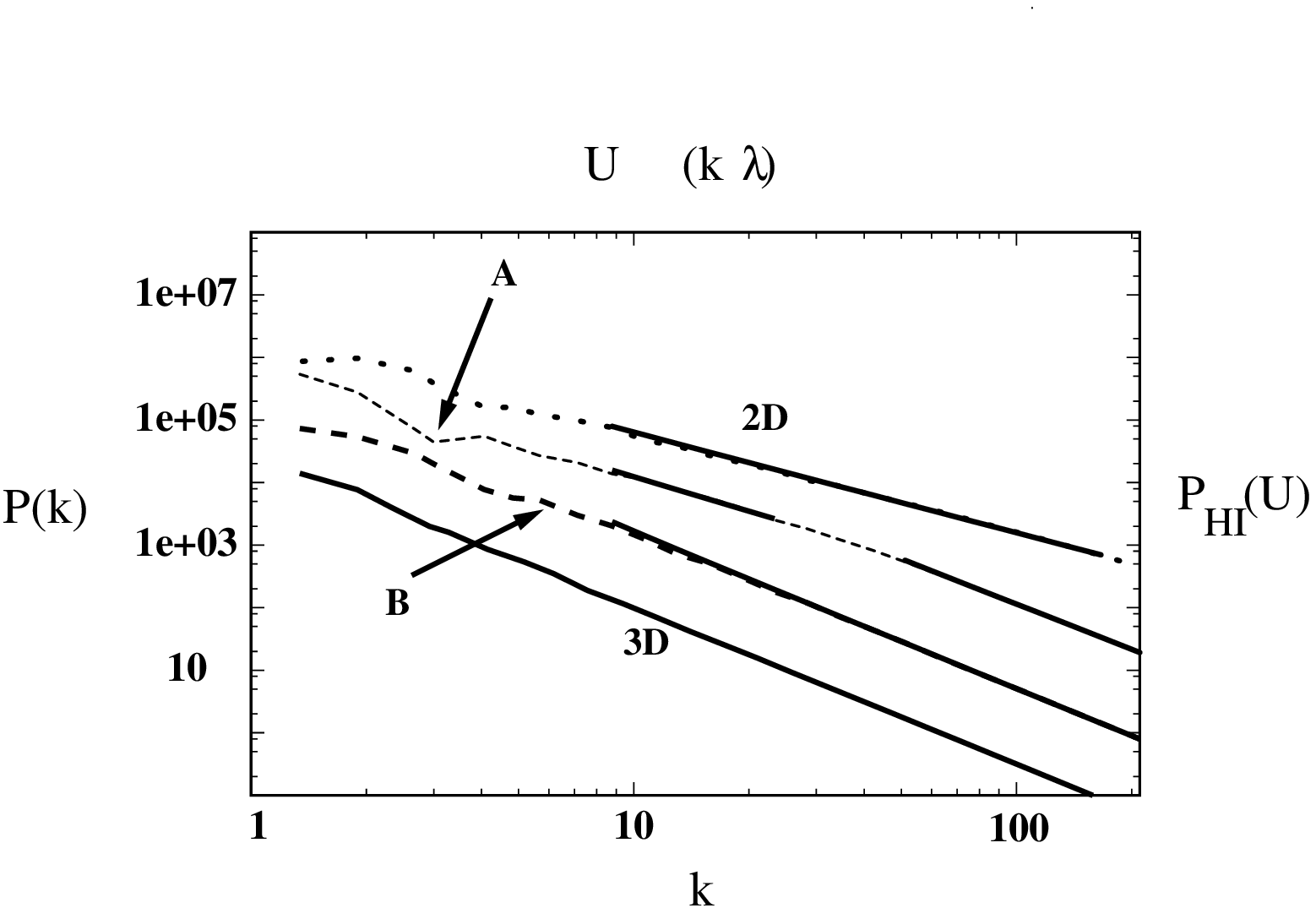, width=3.2in, angle=0}
\end{center}
\caption{The 3D and 2D power spectrum $P(k)$.  
The simulated HI power spectrum $\mathrm{P_{HI}}(U)$
for the thin (A) and thick (B) disk without the radial profile are
also shown.  The $P(k)$ and $k$ values (left and bottom axes) have
been arbitrarily scaled to match the $\mathrm{P_{HI}}(U)$
and $U$ axes (top and right). The power-law fits are shown by  solid
lines.  The different curves have been plotted with arbitrary offsets
to make them distinguishable.
}
\label{fig:sim1}
\end{figure}

\begin{figure}
\begin{center}
\epsfig{file=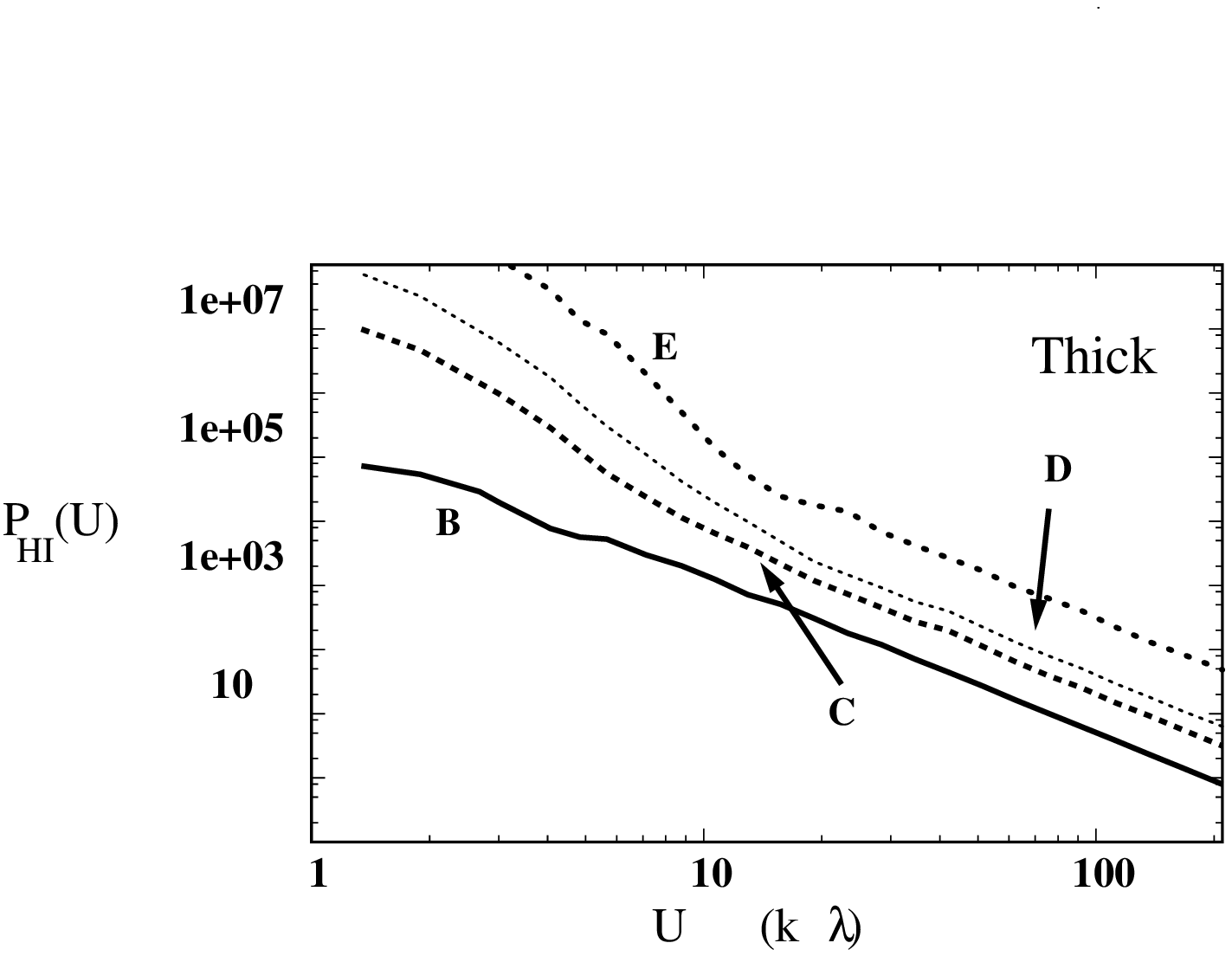,width=3.2in, angle=0}
\end{center}
\caption{The simulated HI power spectrum for the face-on thick disk
  with  (B) no radial profile, and  (C),(D)  with radial profile using 
  $R=0.35,0.7$ respectively.  (E) is same as (C) with the disk
 tilted  at
  $60^{\circ}$.  The different curves have been plotted with arbitrary offsets
to make them distinguishable.
} 
\label{fig:sim2}
\end{figure}

\begin{figure}
\begin{center}
\epsfig{file=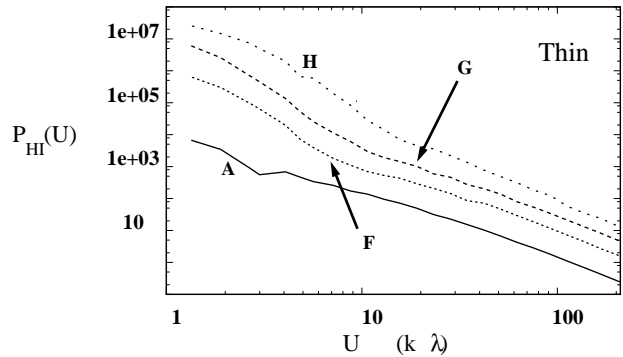,width=3.2in, angle=0}
\end{center}
\caption{The simulated HI power spectrum for the face-on thin disk
  with  (A) no radial profile, and  (F),(G)  with radial profile using 
  $R=0.35,0.7$ respectively.  (H) is same as (F) with the disk  tilted
  at 
  $60^{\circ}$.  The different curves have been plotted with arbitrary
  offsets 
to make them distinguishable. 
} 
\label{fig:sim3}
\end{figure}

We have carried out simulations to asses the impact of the overall
galaxy structure on our estimates of the power spectrum of HI
intensity fluctuations. The starting point is a three dimensional (3D)
$512^3$ cubic mesh, labeled using coordinates $\vr$,  on which we
generate a statistically homogeneous  and isotropic Gaussian random
field $\hat{h}(\vr)$  with power spectrum $P(k)= C k^{\gamma}$. We
have used $\gamma=-2.5$ throughout our simulations, 
the results can be easily generalized to other $\gamma$
values. The 3D power spectrum is shown in  Figure \ref{fig:sim1}.   
For reference we also show the two dimensional (2D) power spectrum of
$\hat{h}(\vr)$ evaluated on a $512^2$ planar section of the cubic
mesh. As expected, the 2D power spectrum also is a power law with
$P(k) \propto  k^{\gamma+1}$.  In all cases we have generated five
independent realizations of the Gaussian random field, and averaged
the power spectrum over these to reduce the statistical
uncertainties.

In our simulations 
we  use the Gaussian random field $\hat{h}(\vr)$ as a  model  for 
the 3D  HI  density fluctuations  that would
arise from homogeneous and isotropic 3D turbulence. The $\hat{h}(\vr)$
values on a 2D section through the cube  serves as a model for the HI
density fluctuations in the limiting situation where we ignore 
the  thickness of the galaxy and treat it as a 2D disk. Note that the
resulting $\hat{h}(\vr)$ is a statistically homogeneous and isotropic
Gaussian random field on the 2D section. We use this to represent the
density fluctuations that would arise from homogeneous and isotropic
2D turbulence.

We next embed a galaxy in the middle of the 3D cube. 
 The overall,  large-scale  structure of 
the galaxy is  introduced  through a function $G(\vr)$ so that the HI 
density at any position is $G(\vr) [h_0 + \hat{h}(\vr)]$. Here $G(\vr)
h_0$ is the  smoothly varying component of the galaxy's 
HI density and $G(\vr)  \hat{h}(\vr)$ is its fluctuating component. 
 It is assumed that the observer's line of sight is  along the
$z$ axis. The HI density    $G(\vr) [h_0 + \hat{h}(\vr)]$ is projected
on the $x-y$ plane. We interpret the projected values 
as  the HI specific intensity $I_{\nu}(\vt)$  in the plane of the
sky (eq. \ref{eq:a1}). The $512^2$ mesh in the $x-y$ plane of our
simulation corresponds  to $4' \times 4'$ on the sky.

We first consider a face-on galaxy, and use $G(\vr)= \exp(-z^2/z_h^2)$.
This incorporates only  the finite thickness of the disk which is
characterized by the scale-height $z_h$ and it ignores  the galaxy's
radial profile  in the plane of the disk. Note that, for $z<z_h$ this
function closely matches the function ${\rm sech}(-z^2/z_h^2)$  
which is also used to model the scale-height profile \citep{BM98}.
For the scale-height, we have used the values  $z_h=4$ and $32$
mesh units 
which we refer to as the ``thin' and ``thick'' disk respectively. 
Figure  \ref{fig:sim1}  shows the  power spectrum 
of the simulated HI specific intensity $I_{\nu}(\vt)$
 for both these cases.  We find that the HI
power spectrum of the thin disk has a slope $-2.5$, same as the 3D
power spectrum, at large $U$.  There is a break
at $U \sim 35 \, {\rm k}\lambda$ which corresponds to $ 1/\pi  z_h$,
and the slope is  $\sim -1.9$ at smaller $U$.  We interpret this
change in the slope in terms of a transition from 3D fluctuations on
scales smaller than the disk thickness (large $U$) 
to 2D fluctuations on scales larger than disk thickness (small $U$).
We expect the slope to approach $-1.5$, the 2D slope,   at very small
$U$.  Note that in our simulation it is not possible to evaluate the
slope at  
very small $U$ values where the sample variance is rather large. 
We find that the HI power spectrum of the thick disk  is well fit by a
single power law $P_{\HI}(U)=A U^{-2.5}$ which has the 3D slope. In
this case we expect the break corresponding to  the transition from 3D
to 2D  at $U \sim 4 {\rm k}\lambda$. This break lies in the  sample
variance dominated region which explains why we do not detect it.

We next incorporate the galaxy's radial profile  using 
\begin{equation}
G(\vr) = \exp \left [-\frac{\sqrt{12}\theta}{\theta_{0}}\right
]\exp(-z^2/z_{h}^2),
\label{eq:galprof}
\end{equation}
where $\theta=\sqrt{x^2+y^2}$  is the radial coordinate  in the
plane of the disk. Here $\theta$ coincides with the angle in the sky
measured from the center of the galaxy.   We have used $\theta_0=1'$
which  corresponds to $128$ mesh units . Note that the radial profile
used in the simulation is  exactly the same as 
the  window function $W(\vt)$ of the exponential model introduced 
in the previous section.   The relative amplitude of $h_0$ and
$\hat{h}(\vr)$ is a free parameter  which  decides the respective
contributions from the  smooth and the fluctuating HI components.   
Using the ratio $R=\sqrt{\langle \hat{h}^2 \rangle}/h_0$
of the rms. value of  $\hat{h}(\vr)$   to $h_0$   
to quantify this, we have carried out simulations for 
$R=0.35$ and $0.7$.  We have also carried out simulations where the
disk is tilted, and the normal to the disk make an angle of
$60^{\circ}$ to the line of sight. 
The  simulated HI power spectra are  shown in 
Figures   \ref{fig:sim2} and \ref{fig:sim3}  for the thick and thin
disks respectively. 

We find that for a face-on disk with $\theta_0=1'$, for both the thick
and thin disks, 
the results of  our simulations are in good agreement with our
analytic  estimate (Section  3) which predicts  that  the effect of
the galaxy's  radial profile is  contained within  a  limited
baseline range  $U \le U_{m}$, which comes out to be 
$U \le U_{m} \sim 3.5  \ \theta_0^{-1}$, for $\alpha = -2.5$ (Figure
  \ref{fig:Um}).
The HI power spectrum is insensitive to the galaxy's  radial  
profile at $U > U_m$, and  the shape of the power spectrum is the same
whether we include the galaxy's radial profile or not. 
We  find that the value of $U_m$ is not very sensitive  
to changes in $R$, the ratio of the fluctuating component to the smooth 
component. For a disk of the same size, the value of $U_m$ increases
when the disk  is tilted.  The image of the disk  tilted  by
$60^{\circ}$ has an anisotropic window function $W(\vt)$ with angular
radius  $1'$ and  $0.5'$ along the major and minor axes respectively.  
For a tilted disk 
 the value of $U_m$ is determined by the smaller of the
 two angular diameters, $0.5'$ in this case. 

In summary, our simulations demonstrate that for baselines $U>U_m$ the
galaxy's radial profile has no effect on the power spectrum of HI
fluctuations. In other words, the power spectrum would be unchanged if
the same fluctuations were present in an uniform disk with  no
radial profile.  Further, a break in the power spectrum at a $U$ value
greater than  $U_m$   is indicative of a transition from 3D to 2D at
the length-scale corresponding to the scale-height $z_h=1/\pi U$. 

Finally, we note that we have ignored velocity fluctuations and 
the galaxy's rotation throughout our simulations. This would 
rearrange the HI emission amongst the  different frequency channels.
Our simulation corresponds to the situation where there is just a
single frequency channel whose width encompasses the entire HI
emission, and the velocities   have no  effect in this
situation. This is justified by the results  presented later in this
paper (Section 6) which   show that, for the galaxies in
our sample,  the slope of  the observed HI power spectrum  does not
depend on the width of the frequency channel.

\section{Method of Analysis}
 The  dwarf  galaxy data used here  is from  Giant Metrewave Radio
 Telescope (GMRT) observations.  
The details can be found in \citet{AJK06} for UGC~4459, 
KDG~52  and  KK~230, \citet{AJK05} for NGC~3741, 
\citet{AJ03} for GR~8, Chengalur et al. 2008 (in preparation) for
AND~IV and  \citet{AJ04} for DDO~210. 
We use Very Large Array(VLA)  archival data,  discussed in
\cite{PASJ07},    for   the spiral galaxy NGC~628.

All the data are reduced in the usual way using  standard tasks in
classic AIPS \footnote{NRAO Astrophysical Image Processing System,   
a commonly used software for radio data processing.}. 
For each galaxy, after calibration the frequency channels with HI
emission were identified and a continuum image was  
made by combining the line free channels. The continuum was hence
subtracted from the data in the $u-v$ plane using  
the AIPS task UVSUB.  

 The number of  channels with HI emission ($n$)  
is  different for  each galaxy. To determine if the HI power spectrum 
changes with the width of the  frequency channel, we combine $N$
successive channels to  obtain $ n/N$ channels and perform the power
spectrum analysis for different values of $N$ in the range $1 \le N
\le n$ . The analysis is initially carried out for $N=1$, and 
unless mentioned otherwise the results refer to this value. 

We measure  $\hat{P}_{\rm  HI}(U)$ by  correlating  every visibility
$V_{\nu}(\vU)$ with all other visibilities  $V_{\nu}(\vU + \Delta
\vU)$  within a disk $\mid \Delta \vU \mid \ll \theta_{0}^{-1}$ and
then average over different $\vU$ directions. Note that correlation of
a visibility with itself is excluded.
To increase the signal
to noise ratio we further average the correlations in  bins of $U$ and
over all  frequency channels with HI emission.

The measured visibility correlation estimator (eq. \ref{eq:corrn}) is
the convolution of the actual HI power spectrum with a window
function. Here we do not attempt to estimate this window function by
fitting  the  observed HI image and then use this to deconvolve the
power spectrum. Our approach is to estimate a baseline  $U_m$, such
that for $U \ge U_m$  the effect of the convolution can be ignored.
The measured visibility estimator is  proportional to the HI power
spectrum (eq. \ref{eq:corra}) at baselines $U \ge U_m$. 
We estimate the HI power spectrum using  only this range ($U \ge U_m$)
where the measured visibility  correlation estimator may be directly
interpreted as the HI power spectrum.

Assuming an exponential window function, the value of $U_m$ depends on
the parameters $\theta_0$ and $\alpha$.  Note that we do not attempt
to actually fit an exponential  to the observed overall HI
distribution and thereby determine $\theta_0$. 
 The angular radius $\theta_0$
of the galaxies  in our sample is set to half the angular extent
listed in  Table~\ref{tab:profile}. 
The galaxies are typically ellipses and
not circular disks as assumed in Section 3. Guided by our simulations, 
we have used the smaller
of the two angular extents  in our analysis. As $U_m\propto
\theta_0^{-1}$, choosing the smaller value  gives a 
conservative estimate of $U_m$. The upper limit $U_u$ where we 
have a reliable estimate of the HI power spectrum is  determined
by the requirement that the  real part of $\hat{\rm
  P}_{\HI}(U)$ should be more than its  imaginary part and also the 
noise in  the line free channels. 

The procedure that we adopt to determine the best fit power law to the
HI power spectrum is as follows.  We first visually identify a
baseline $U_m$ beyond which ($U \ge U_m$) the visibility correlation
estimator appears to be a power law in  $U$.  We then use the range
$U_m \le U \le U_u$ to  fit a power law  $\hat{\rm
  P}_{\HI}(U)=A~U^{\alpha}$. The best fit $\alpha$ obtained by this
fitting procedure  is used to obtain a revised  estimate for $U_m$
through Figure \ref{fig:Um}.  The value of $U_m$ in this figure are
for $\theta_0=1'$, we use the scaling $U_m \propto \theta_o^{-1}$
(as mentioned above, $\theta_{0}^{-1}$ is taken to be the smaller
value of the two angular extents for a particular galaxy) to
obtain the $U_m$ value that is appropriate for the galaxy.  The 
power law fitting is repeated with the revised estimate of 
$U_m$.  We iterate this procedure a few times till it converges. 
At each iteration, the best fit  $A$ and $\alpha$ were determined
through a $\chi^2$    minimization. To test whether the impact of the
window function  is actually small,  we have convolved the best fit
power law  with $\mid  \tilde{W}_{\nu}(U)\mid^2$. The convolved power
spectra were visually 
inspected to asses the deviations from the power law. The goodness of
fit to the data was also estimated by calculating $\chi^2$ for the
convolved power spectrum.  The final fit is accepted only after
ensuring that the effect of the convolution can actually be ignored.

The values  $D/U_m$ and $D/U_u$, where $D$ is the distance
to the galaxy,  give an  estimate of the range of length-scales over
which the  power-law fit holds.  In addition to the other parameters
of the power law fit, these values are also shown in Table~2.

\section{Result and Conclusion}
\label{sec:result}
\begin{figure*}
\begin{center}
\epsfig{file=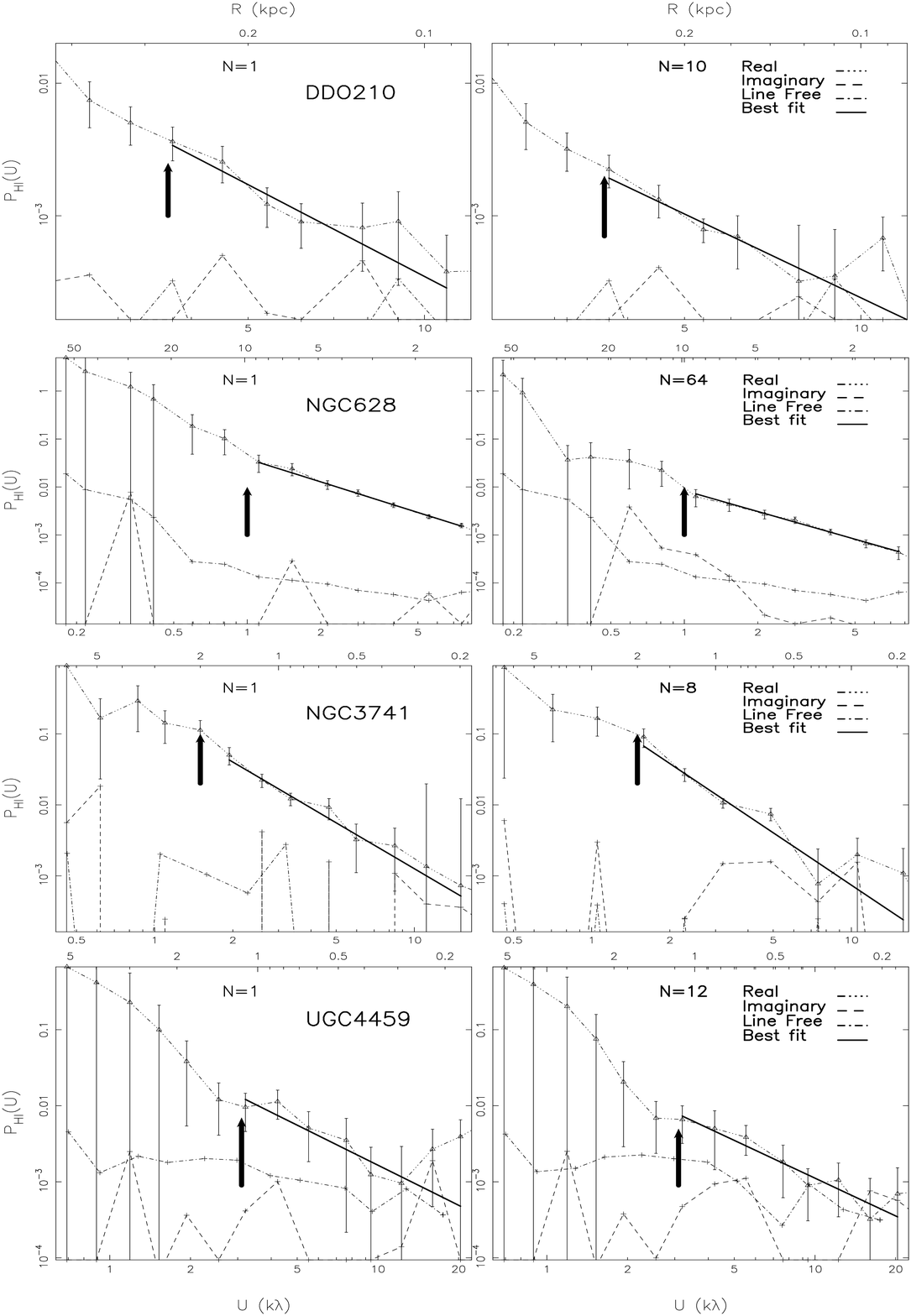,width=5.6in, angle=0}
\end{center}
\caption{Power spectrum of the galaxies DDO~210, NGC~628, NGC~3741 and
  UGC~4459. The real and imaginary parts   of $\hat{P}_{\rm  HI}(U)$
estimated after averaging $N$ channels with HI emission, and the real
  part from $N$ line-free channels are shown together for $N=1$  (left
  panel) and $N=n$ (right panel). The error-bars are for the real part
  from channels with HI emission.    The best fit power law is shown
  in bold.  In each case $U_{m}$ is marked with  a  bold-faced arrow
  and   the fit is restricted to $U > U_m$ where the
  effect of the convolution with the window function can be
  ignored.}
\label{fig:mom0}
\end{figure*}

\begin{figure*}
\begin{center}
\epsfig{file=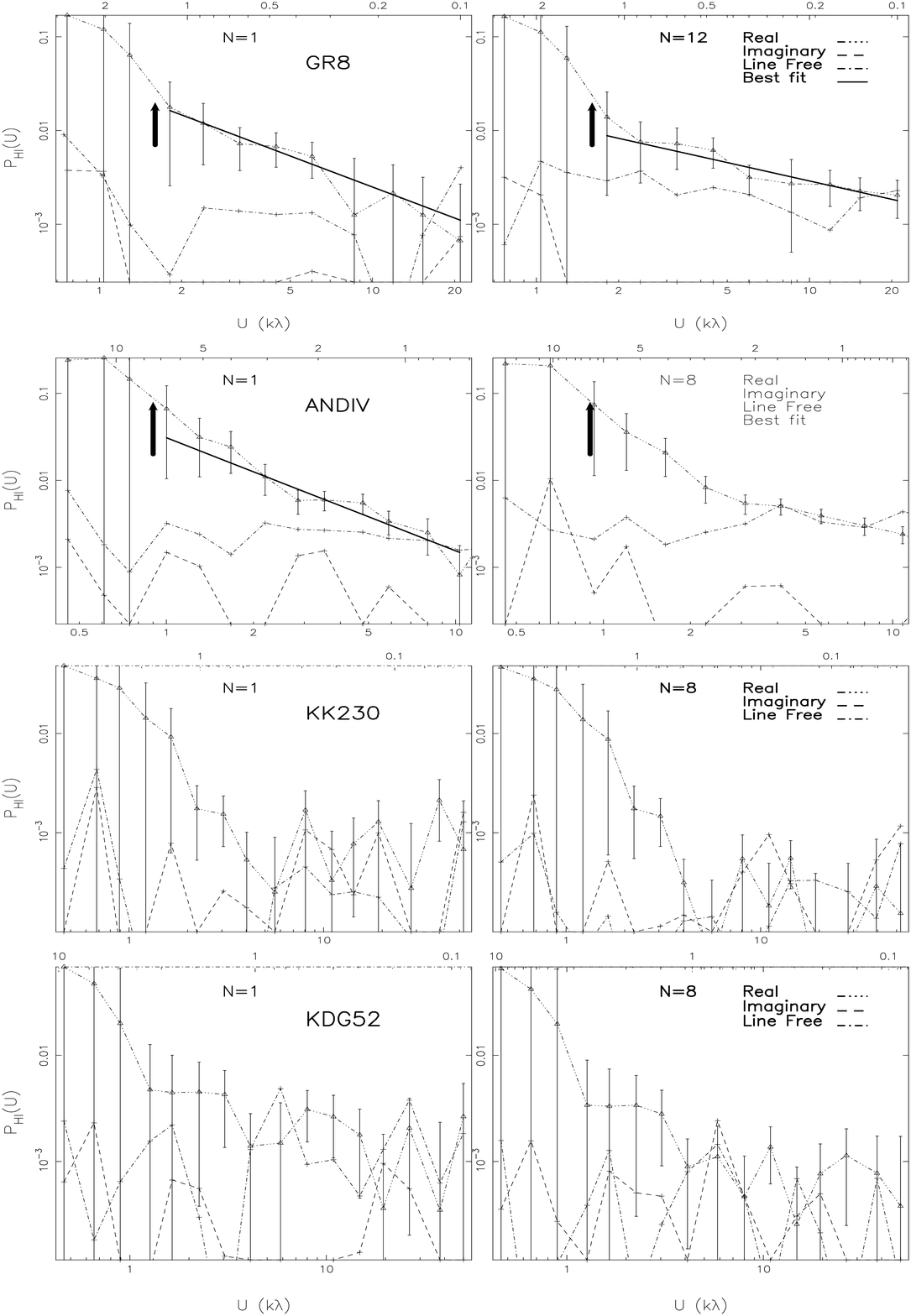,width=5.6in, angle=0}
\end{center}
\caption{Power spectrum of the galaxies  GR~8, AND~IV, KK~230 and
  M81DWA.    The real and imaginary parts   of $\hat{P}_{\rm  HI}(U)$ 
estimated after averaging $N$ channels with HI emission, and the real
  part from $N$ line-free channels are shown together for $N=1$  (left
  panel) and $N=n$ (right panel). The error-bars are for the real part
  from channels with HI emission.    The best fit power law is shown
  in bold only where such a fit is possible. For those,    $U_{m}$ is
  marked with  a  bold-faced arrow   and   the fit is restricted to $U
  > U_m$ where the   effect of the convolution with the window
  function can be   ignored. }
\label{fig:mom01}
\end{figure*}

Figure~\ref{fig:mom0}  and  \ref{fig:mom01} show the results of our
analysis. The results are summarized in Table~\ref{tab:result}.
  We have detected the HI power spectrum of  
the galaxies  DDO~210, NGC~628, NGC~3741, UGC~4459, GR~8
and AND~IV. For these galaxies we find a range of baselines $U$  where the
real part of $P_{HI}(U)$ estimated from the channels with HI emission
is larger than the imaginary part estimated  from the same channels
as well as  the real  part estimated from the line free channels. 
This is not true  for KK~230 and KDG~52 (and also for AND~IV with $N=n$),  for these galaxies  
all three curves (Figure \ref{fig:mom01})
lie within the $1-\sigma$ errors-bars with very small offset and the interpretation is not
straight-forward.   For  the subsequent analysis of all the 
galaxies we use only the real part of  $P_{HI}(U)$ estimated from the
channels with HI emission.  

For all the galaxies  we have tried to fit the observed $P_{HI}(U)$
 with a  power law  $P_{\rm HI}(U)=A U^{\alpha}$.  We find that this 
provides  a good fit with a reasonable  $\chi^{2}/\nu$  for 
DDO~210, NGC~628, NGC~3741, UGC~4459, GR~8 and AND~IV.  The presence
 of a scale invariant, power law power spectrum indicates that
 turbulence  is operational in the ISM of these galaxies. 
The range of  length-scales for the power law  fit differs from galaxy
 to galaxy and in total  it covers  $100$pc to $8$Kpc. The details are
 summarized in 
 Table~\ref{tab:result}. 

\begin{table*}
\centering
\begin{tabular}{|l|c|c|c|c|c|c|c|c|c|}
\hline
Galaxies & DDO~210 & NGC~628 & NGC~3741 & UGC~4459 & GR~8 & AND~IV & KK~230 & KDG~52 \\
\hline \hline
(1 a) $N$    &   1   &   1     &    1   &   1    &  1     &   1    &    1  & 1  \\
(2 a) velocity width (km s$^{-1}$)& 1.65&1.29&1.65&1.65&1.65&1.65&1.65&1.65\\
\hline
(3 a) $\alpha$ & $-2.3\pm0.6$&$-1.6\pm0.2$&$-2.2\pm0.4$&$-1.8\pm0.6$&$-1.1\pm0.4$&$-1.3\pm0.3$&-&-\\
(4 a) $U_m$ - $U_u$ (k$\lambda$) & $3.7-13.0$&$1.0-10.0$&$1.6-16.0$&$3.1-22.0$&$1.6-23.0$&$0.6-6.7$&-&-\\
(5 a) $D/U_u$ - $D/U_m$ (kpc)& $0.06-0.27$&$0.8-8.0$&$0.19-1.9$&$0.18-1.16$&$0.1-1.3$&$0.56-6.2$&-&-\\
(6 a) $\chi^{2}/\nu$& 0.3&0.2&0.2&0.7&0.6&0.4&-&-\\
\hline \hline
(1 b) $N$     & 10&64&8&12&8&16&8&8\\
(2 b) velocity width (km s$^{-1}$)& 16.5&82.56&13.2&19.8&13.2&26.4&19.8&19.8\\
\hline
(3 b) $\alpha$ & $-2.1\pm0.6$&$-1.5\pm0.2$&$-2.5\pm0.4$&$-1.7\pm0.4$&$-0.7\pm0.3$&-&-&-\\  
(4 b) $U_m$ - $U_u$ (k$\lambda$) & $3.7-13.0$&$1.0-10.0$&
$1.6-16.0$&$3.1-22.0$&$1.6-23.0$&-&-&-\\ 
(6 a) $D/U_u$ - $D/U_m$ (kpc)&
$0.06-0.27$&$0.8-8.0$&$0.19-1.9$&$0.18-1.16$&$0.1-1.3$& - &-&-\\
(6 b) $\chi^{2}/\nu$& 0.4&0.2&1.3&0.3&0.2&-&-&-\\
\hline \hline
(7)   $\zeta_{2}$& $0\pm1.2$ & $0\pm0.4$ & $0\pm0.8$& $0\pm1.2$& $0\pm0.8$& $0\pm0.6$ &- &- \\
\hline \hline  
\end{tabular}
\caption{The results for  the 8 galaxies in our sample.  Rows 1-6
  gives (1) $N$  the number of channels  averaged over, (2) the 
  corresponding  velocity width 
  (3) $\alpha$ the best fit slope for the HI power spectrum, 
  (4) $U$ range over which the power law fit is valid,  (5) length-scales
  over which the power law fit is valid  and (6) the
  goodness  of fit $\chi^2$ per degree of freedom. Row (7) Possible
  limits for the spectral slope of the velocity structure function.} 
\label{tab:result}
\end{table*}

Both HI density fluctuations as well as spatial fluctuations in the
velocity of the HI gas contribute to fluctuations in the HI specific
intensity.
Considering a turbulent ISM, \citet{LP00} have shown that 
it is possible to disentangle these two contributions  by studying
the behavior of the HI power spectrum as the thickness of the
frequency  channel is varied. If the observed HI power spectrum is 
due to the gas velocities, the slope of the  power spectrum 
is predicted to decrease with increasing thickness of the 
frequency  channel.   To test this we have   repeated the power
spectrum analysis 
increasing  the channel thickness $N$ from $N=1$ to $N=n$.   In
addition to $N=1$,   Figures \ref{fig:mom0}  and  \ref{fig:mom01},
and  Table~\ref{tab:result}  also show the  results for $N=n$
  where the channel thickness spans the entire frequency range that
  has   significant  HI emission.   We do  not find a
  significant change in the slope of the power spectrum   for any of
  the galaxies.    
Since for all the galaxies   the thickest channel  is considerably
larger  than the velocity dispersion, we conclude that    
 the HI power  spectrum is  purely due to  density
fluctuations and not gas velocities. The fact that the slope does not
change with channel thickness can be used to  
constrain the value of $\zeta_{2}$,  the  slope of the
velocity structure function (eg. \citealt{LP00}).  
The $\zeta_2$ values   are tabulated in  Table~\ref{tab:result}.

The galaxies in our sample have slope $\alpha$ ranging from $-2.6$ 
to $-1.1$.  The two galaxies DDO~210 and NGC~3741 have slope $\sim
-2.5$, while the slope is  $\sim -1.5$ for  NGC~628, UGC~4459 and
AND~IV,   and $-1.1$ for   GR~8.   We have proposed a possible
explanation  for this dichotomy in the  values of $\alpha$ in
\citet{PASJ07}.  This was based on the fact that DDO~210, where the
power spectrum was measured across length-scales $100-500 \, {\rm
  pc}$, had a slope  of $-2.6$  while NGC~628, a nearly face-on galaxy
where  the power spectrum was measured across length-scales $0.8-8 \,
{\rm   kpc}$ had a slope of $-1.6$. We have interpreted the former as
three 
dimensional (3D) turbulence operational at small scales whereas the
latter was interpreted as two dimensional (2D) turbulence in the plane
of the galactic disk. For a nearly face-on disk galaxy we expect the
transition from 2D to 3D turbulence to be seen at a length-scale
corresponding to the scale height of the galaxy. 
Continuing with this interpretation implies that we have also measured  
3D turbulence in NGC~3741,  and 2D turbulence in   UGC~4459,  GR~8 and
AND~IV.

The power spectrum analysis can be used to determine the scale-height 
of nearly face-on galaxies  \citep{EK01,PK01}. 
 \citet{EK01} have applied this method 
to  estimate the scale height of LMC. In a recent study \citep{PASJ07}
we  have  placed an upper limit of $800 \, {\rm pc}$ for the
scale-height of NGC~628. Our simulations (Section 4) show that 
we expect the slope of the HI power spectrum to change 
 corresponding to a transition from 3D to 2D turbulence at the
scale-height.  The baseline
 $U$  at which this break occurs can be used to  estimate  the 
 scale-height $D/\pi U$, where $D$ is the distance to the galaxy.

In addition to NGC~628, our  sample contains   3 more nearly face-on  
galaxies with $i_{\rm HI} < 30^{\circ}$, namely  
 DDO~210, UGC~4459 and GR~8. DDO~210 has 3D turbulence across the 
baselines $2.0-10.0 \ {\rm k \lambda}$. The absence of a break in the
power spectrum  places a lower limit of $160$   pc 
on the  scale-height.  The galaxies NGC~628, 
UGC~4459 and GR~8  exhibit 2D turbulence for the entire $U$ range 
(Table~2) in the  measured HI power spectrum. This 
imposes  the upper limit of $320 \ {\rm pc}$, $51 \ {\rm pc}$
and $30 \ {\rm  pc}$ respectively on  the
scale-height.  Note that the scale-height of $320 \ {\rm pc}$ for
NGC~628  differs from our earlier  estimate \citep{PASJ07} because of
the extra factor of $\pi$ indicated by our simulations.

\begin{figure*}
\begin{center}
\epsfig{file=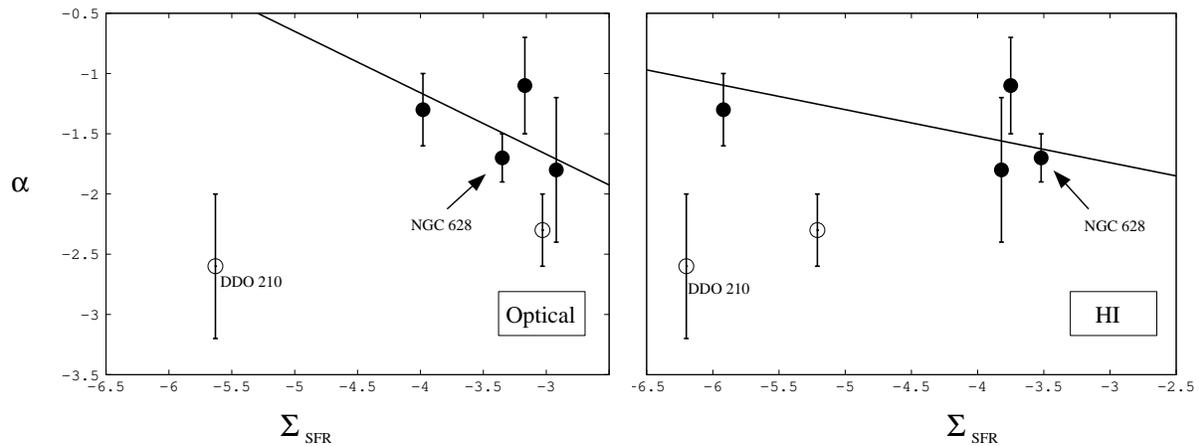, angle=0, width=6.2in}
\end{center}
\caption{The slope $\alpha$ of the HI power spectrum plotted against
  $\Sigma_{{\rm SFR}}$, the   SFR per unit area. The area has been 
determined from optical images and HI images in the left and right
panels respectively. The  galaxies with  3D and 2D
turbulence are shown using empty and filled circles respectively. 
Note that the SFR for DDO~210, marked in the figure, is only 
 an upper limit. The data for  NGC~628, which is a spiral galaxy, is
 marked with an arrow.  The straight lines show  the linear
 correlation that we find (discussed in the text)   between the
 slope of the  HI power spectrum and  the SFR per unit area.   
} 
\label{fig:SFR}
\end{figure*}

The energy input from star formation is believed to be a major driving
force for the turbulence in the ISM \citep{ES04I}. Hence, it is very
interesting to check whether the slope of the power spectrum of
intensity fluctuations in these galaxies has any correlation with the
SFR. Any correlation, if present, will provide an insight into the
processes driving turbulence in the  ISM.

In a recent study \citet{WED05} calculate the SFR per unit area for  9
irregular 
galaxies  and investigate the correlation with the 
V-band and  H$\alpha$ power spectra.   The length-scales they  probe
are  $10-400$ pc. In  H$\alpha$ they  find that  the power spectra
becomes steeper as the SFR per unit area  increases. However  they do
not find  any   correlation in the V-band.  In this paper we probe
 length-scales $100$ pc to $3.5$ kpc. 
We test for a   possible correlation between the slope 
of the power spectrum  and the  SFR per unit area and per unit HI mass  for  the 5 dwarf
galaxies in our sample (Figure~\ref{fig:SFR}).
 We report results using  the area estimated
from  both,  the HI disk and the optical disk.  
 The data for the SFR, angular extent,  inclination in HI and HI mass   for
 these galaxies are from \citet{AJK05} and  \citet{AJK08}. The SFR is
 determined from H$\alpha$ emission,  and the 
 angular extent in HI is determined from  the Moment0 maps at a column
 density of  $10^{19}$ atoms~cm$^{-2}$.  For the angular extent and
 inclination  of the  optical disk   we use the   parameters from
 \citet{Z00}, \citet{AJK06}  and \citet{S08}.
The  linear correlation coefficient   is found to have  values $0.35$
and  $0.34$ for the HI and optical disks respectively, indicating the
absence of any correlation between the slope of the power spectrum and
the SFR per unit area. We have similar result for SFR per unit HI mass
for these 5 galaxies. However, we  note a few points that should
be taken into 
consideration when interpreting this conclusion. The H$\alpha$
emission is  not a good tracer of SFR in low mass galaxies due to
stochastic star formation   \citep{KK07, L07}.  Also, the length-scales
across which the  power spectrum  has been estimated for  the
 dwarf galaxies in our sample are substantially larger than the
 optical disk where  star formation occurs.

Turbulence can also possibly be related to  different parameters
of the  galaxy. We test  for correlations between the slope of the HI
power spectrum and   the following parameters: total HI
mass,  HI mass to light ratio, total dynamical mass, total baryonic
mass,  gas fraction, baryon fraction. We use the estimates of these
observable  quantities from \citet{AJK07}. In all the cases the linear
correlation coefficient is found to lie between $-0.5$  and $0.5$,
indicating the absence  of correlation.  

In the above analysis we  have considered galaxies with both 2D and 3D
turbulence 
taken together, which in turn can suppress the correlations.   
Hence, we further investigated the correlations by considering the 
 4 galaxies in our sample (including NGC~628) with 2D
turbulence.  In
this case the correlation coefficient comes out to be -0.70 and -0.74
for SFR per unit area of optical and HI disks respectively. This
indicates a strong correlation and the result is similar to that of
\citet{WED05}, namely that the  surface density of star formation rate
is larger  for galaxies with steeper HI power spectrum. These findings
indicate  a possible link between star formation and 
the nature of turbulence in the ISM. However, it would not be 
realistic  to speculate on a possible cause and effect relation
between these two. It is quite possible   that the observed correlation
is an outcome of  extraneous factors which influence both.

Finally, we note that the total number of galaxies in all of our analysis is
rather small  for a statistical conclusion. A larger galaxy sample  is
required for a better  understanding of the  generic  features,  if
any,  of turbulence  in  the ISM of faint dwarf  galaxies.

\section*{Acknowledgments}

P.D. is thank full to Sk. Saiyad Ali, Kanan Datta, Prakash Sarkar,
Tapomoy Guha Sarkar, Wasim Raja and Yogesh Maan for use full
discussions. P.D. would like to acknowledge SRIC, IIT, Kharagpur for
providing financial support. S.B. would like to acknowledge financial
support from 
BRNS, DAE through the project 2007/37/11/BRNS/357. Data presented in
this paper were obtained from GMRT (operated by the National Centre
for Radio Astrophysics of the Tata Institute of Fundamental Research)
and NRAO VLA.

\end{document}